\documentclass[journal]{IEEEtran}
\usepackage{orcidlink}
\usepackage{cite}
\usepackage{amsmath,amssymb,amsfonts}
\usepackage{algorithm,algorithmic}

\usepackage{graphicx}
\usepackage{textcomp}
\usepackage{hyperref}
\usepackage{bookmark}
\usepackage{booktabs}
\usepackage{float}
\usepackage{xcolor}
\usepackage{soul}
\def\BibTeX{{\rm B\kern-.05em{\sc i\kern-.025em b}\kern-.08em
    T\kern-.1667em\lower.7ex\hb\ox{E}\kern-.125emX}}
\begin{document}
\title{{\normalsize This work has been submitted for possible publication.\vspace{-0.6cm} Copyright may be transferred without notice, after which this version may no longer be accessible.\vspace{0.8cm}}\\
An IoT Framework for Building Energy Optimization Using Machine Learning-based MPC
}

\author{Aryan~Morteza\,\orcidlink{0009-0004-7457-3124},
	    Hosein~K.~Nazari\,\orcidlink{0000-0003-1632-4782},
	    Peyman~Pahlevani\,\orcidlink{0000-0001-5918-7250}
}

\maketitle
\begin{abstract}
This study proposes a machine learning-based Model Predictive Control (MPC) approach for controlling Air Handling Unit (AHU) systems by employing an Internet of Things (IoT) framework. The proposed framework utilizes an Artificial Neural Network (ANN) to provide dynamic-linear thermal model parameters considering building information and disturbances in real time, thereby facilitating the practical MPC of the AHU system. The proposed framework allows users to establish new setpoints for a closed-loop control system, enabling customization of the thermal environment to meet individual needs with minimal use of the AHU. The experimental results demonstrate the cost benefits of the proposed machine-learning-based MPC-IoT framework, achieving a 57.59\% reduction in electricity consumption compared with a clock-based manual controller while maintaining a high level of user satisfaction. The proposed framework offers remarkable flexibility and effectiveness, even in legacy systems with limited building information, making it a pragmatic and valuable solution for enhancing the energy efficiency and user comfort in pre-existing structures.
\end{abstract}

\section{Introduction}
\label{sec:introduction}

Residential, public, and commercial buildings account for 47\%, 49\%, and 12\% of the energy use in electricity, natural gas, and petroleum products, respectively \cite{ramin2020optimum}. Implementing an effective control system for heating, ventilation, and air conditioning (HVAC) enhances energy consumption efficiency, elevates residents' satisfaction, and reduces their carbon footprints. Most studies have focused on high-tech buildings with advanced HVAC controllers, thus limiting the benefits to newer structures. Older commercial buildings and smaller residential areas cannot afford these systems. Studies have demonstrated that incorporating basic functionalities in the initial phase requires an average cost of \$2.50 per square foot \cite{rawal2016costs}. This group of buildings embodies an untapped reservoir of energy-saving potential.

Typically, an HVAC system undergoes a two-phase process: first, comprehend the building model and subsequently integrate this model into the controller to facilitate effective decision-making. In the first phase, several studies employed building envelope information to comprehend the building model, which included the geometric and insulation properties of the building walls, roof, windows, and doors. Various models, such as linear \cite{aswani2011reducing}, bilinear \cite{sturzenegger2015model}, and nonlinear \cite{aswani2012energy}, have utilized building envelope information in their respective studies. Acquiring building envelope information is a resource-intensive process that demands individual attention from each building, considering factors such as geographical location and specific physical attributes. In response to this challenge, alternative approaches proposed in existing studies advocate learning building models through Machine Learning (ML)-based methods. Several studies, including \cite{ferreira2012neural, aswani2012energy, bunning2020experimental} utilized historical data collected from sensors—informational repositories to capture a building's behavior to construct a building model using ML techniques. These techniques can be broadly categorized as static and adaptive modes. Buildings that, function as transient systems with time-dependent parameters, experience variations in properties such as heat transfer owing to factors such as local weather conditions, thermal mass, and surface characteristics \cite{yang2019adaptive}. Failure to adapt models to these changes can result in loss of accuracy in capturing the time-dependent characteristics of a building. Implementing adaptive models for building thermal systems, which evolve dynamically, enhances modeling precision and Model Predictive Control (MPC) performance. The adaptive model presented in \cite{zhang2019iot} addresses information-gathering in older buildings using a cost-effective approach. However, the evaluation only covers outcome prediction, lacking insights into the real-world impact, potential energy savings, and practical implementation. This discussion overlooks the details of utilizing the proposed method for controllers, and its compatibility with different types of controllers.

Thermal models were employed to manage the controllers in the second phase of the HVAC system. Modern buildings typically utilize controllers that receive specific setpoints, whereas many older buildings rely on controllers that are limited to binary decision-making  (ON/OFF) processes. Advanced control techniques, such as Model Predictive Control (MPC) leverage building thermal models to generate precise commands for different types of controllers \cite{stoffel2023evaluation}. Many studies, including \cite{sturzenegger2015model,yang2020model}, have focused on developing MPC for setpoint-based controllers to optimize decision-making. By contrast, actuators employing conventional low level controllers (e.g., PI, PID, or embedded controllers) cannot harness the generated setpoint-based outputs. Numerous studies, such as \cite{aswani2011reducing} and \cite{carli2020iot}, have offered insights into the utilization of MPC in conjunction with binary-based controllers. Notably optimizing a system using a binary-based controller poses significant challenges. Although the results of these studies demonstrate the advantages of such systems, none of them have considered adaptive learning models for thermal models. Consequently, there is substantial potential for further enhancement of these methods.

\begin{figure*}[t]
\centering
\includegraphics[width=.70\textwidth]{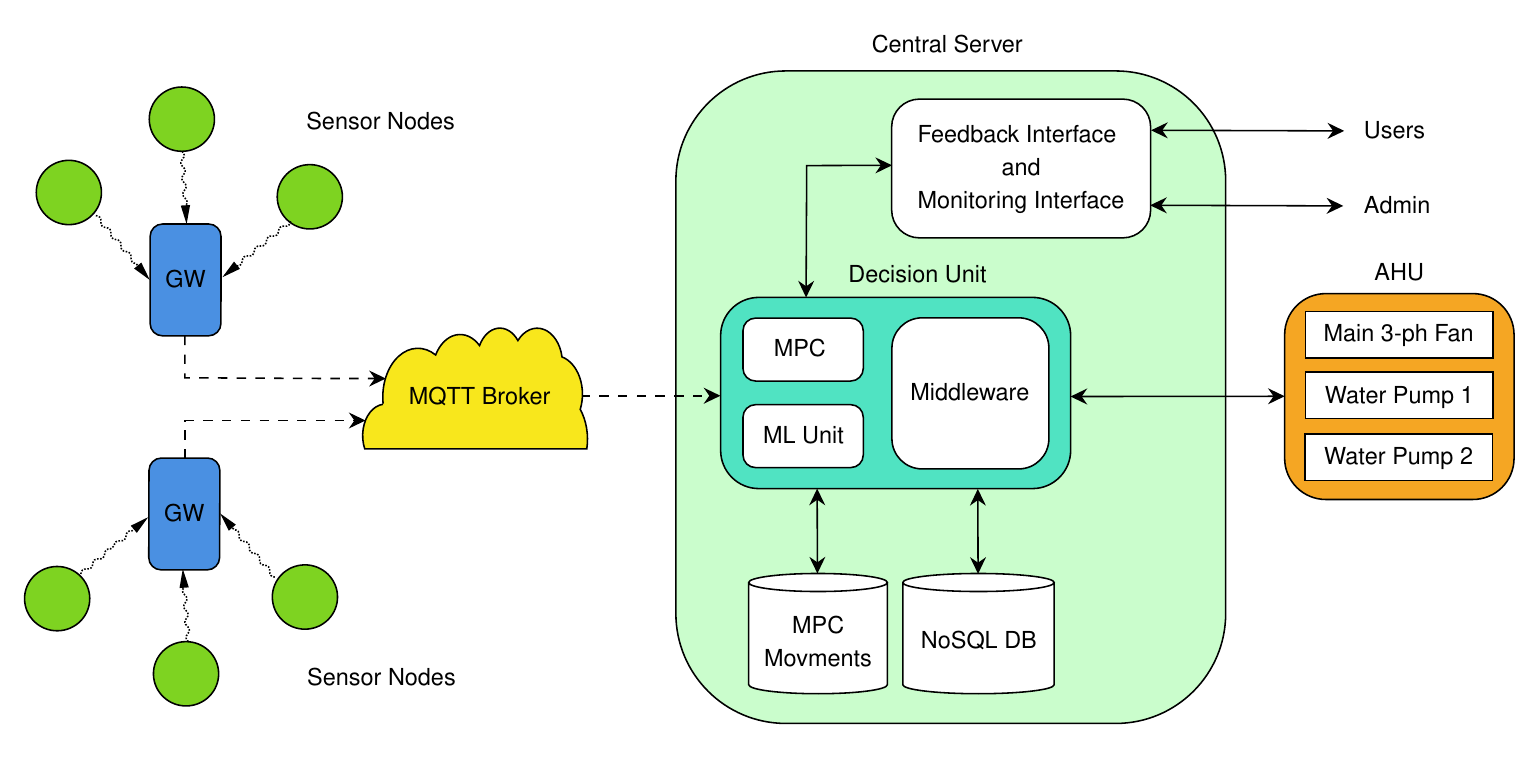}%
\caption{The experimental setup involved collecting stream data at five-minute intervals using Raspberry Pis. The collected data were then transmitted to the central server via the MQTT protocol, which enabled the AHU system to make decisions based on the occupants/administrators given a setpoint.}
\label{IoT-solution}
\end{figure*}

Although extensive research has been conducted on modern building HVAC systems, there is a noticeable lack of attention directed toward older buildings with fewer capabilities and resources. This study addresses this gap by focusing on buildings that employ ML-based techniques for thermal models, thereby eliminating the requirement for expensive envelope information. 
Additionally, we explored systems equipped with binary decision-making controllers. This type of controller adds complexity to tasks owing to the limited on-off operations and the need for more precise building models.
This paper presents an IoT-based Framework that employs a refined version of the ML-based method inspired by \cite{zhang2019iot}. Our revised modeling method improves the sample point usage and incorporates additional inputs, including weather conditions, such as wind speed and solar radiation, to enhance the building model.
In addition, we integrated parameter extraction to ensure compatibility with the binary controllers for the output of the model. In addition to HVAC system modeling and control, our study employed low-cost Internet of Thing (IoT) embedded systems featuring a scalable framework using a LoRa-based Wireless Sensor Network (WSN). Simultaneously addressing user satisfaction and energy reduction, the framework includes a real-time User Interface (UI) for monitoring and analyzing thermal data, setpoints, and energy consumption. The overall system architecture is illustrated in Figure \ref{IoT-solution}.
To determine the effectiveness of our proposed approach, we conducted a thorough evaluation of the proposed system over 126 days. This large-scale experiment involved the deployment of 24 sensor nodes within a building comprising of 24 rooms. The building, oriented in a north-south direction, features an HVAC system located on the first floor. Figure \ref{real and blueprint} shows a blueprint depicting one side of the building floor. Our results show a significant reduction of 57.59\% in electricity consumption compared with a clock-based manual controller. In addition, our building modeling techniques outperform the method presented in \cite{zhang2019iot} by providing a more accurate prediction, with an average reported scaled Mean Absolute Error (MAE) of 10\% compared to 20\%. Furthermore, our model exhibited a notable 30\% increase in explained variance.

The remainder of this paper is organized as follows. A literature review is presented in Section \ref{section:2}. The different components of the testbed are described in Section \ref{section:3}. Section \ref{section:4} details the ML structure used to construct the building models. The MPC-based control algorithm for HVAC systems and Section \ref{section:5} and an analysis of the results are provided in Section \ref{section:6}. Section \ref{section:7} provides a summary of the main points discussed and suggests potential future work.

\begin{table*}
\caption{Summary and comparison of smart building research using the MPC.}
 \begin{tabular*}{\textwidth}{c @{\extracolsep{\fill}}llllll}
  \toprule
  \textbf{Study} & \textbf{Place of implementation}& \textbf{Actuator(s)}& \textbf{Duration} & \textbf{MPC model}& \textbf{Comm. equipment (Sensor Network)}\\ 
  \midrule
  
\cite{clarke2002simulation} & 1 empty room & Radiator & 3 hours  & ESP-r & BEMS (Not specified) \\

\cite{henze2005experimental} & 2 empty rooms & Chillers, AHU, and Ice storage & 4 days & TRNSYS & BAS (Not specified)\\

\cite{liao2009inferential} & Commercial building & Boiler supplying passive radiator & 40 days  & Linear & NM\\

\cite{aswani2011reducing}  & 1 room & Air conditioner & 24 hours & Linear (ODE) & BRITE (Wireless)\\

\cite{schuss2011empirical} & 2 and 3 occupied room & Windows and veils & 60 days  & HAMbase-Radiance & NM \\

\cite{vsiroky2011experimental} & 8 floors building & Floor heating & 90 days & Linear & BAS (Not specified) \\

\cite{ma2011model} & University building & Chillers and cooling towers & 10 days  & Nonlinear & NM \\

\cite{aswani2012energy} & 7 floors building & HVAC & 8 days & Nonlinear & BRITE (Not specified) \\

\cite{ferreira2012neural} & 4 occupied room & AHU & 3 days & ANN & LonWorks bus (Wireless) \\

\cite{dong2014real} & Solar house & HVAC & 67 days & Nonlinear & NM (Hybrid) \\

\cite{castilla2014thermal} & 1 room  & Fan coil & 12 hours & Nonlinear & NM \\

\cite{bengea2014implementation} & Research building & AHU & 21 days & Nonlinear & BMS (Hybrid)\\

\cite{sturzenegger2015model} & Offices building  & TABs, veils and air conditioner & 203 days & Bilinear & BAS (Wireless)  \\

\cite{de2016practical} & 2 floors building & Boiler and heat pumps & 12 days  & Nonlinear & BECMS (Not specified) \\

\cite{afram2017supervisory} & 3 floors building & HVAC & 25 days & Linear & NI-CFS (Wired) \\

\cite{fiorentini2017hybrid} & Solar house & Air conditioner  & 2 days & Nonlinear &  BMCS (Not specified)  \\

\cite{chen2016occupant} & Chamber & HVAC (Heater and chiller) & 36 hours & Wiener model  & HOBO U12 (Wired)  \\

\cite{sangi2019real} & 2 floors building  & HVAC & 9 hours  & Linear & BSM-BACNet (Wired) \\

\cite{joe2019model}  & 3 Labs  & AHU and radiant floor  & 20 days  & Linear  & BMS-Niagara/AX (Wireless) \\

\cite{yang2020model} & Office and lecture theater & AC and Mechanical Ventilation & 20 days & NARX RNN & BMS-BAC (Not specified) \\

\cite{bunning2020experimental}   & Apartment & Radiator valves & 12 days & ANN & NM \\

\cite{drgovna2020cloud} & Offices building  & AHU and TABS  & 14 days & Nonlinear &  Cloud-based SCADA (Not specified)\\

\cite{saloux2021practical} & Institutional building    & Electrical and gas boiler   &  118 days   & Linear &  BAS (Not specified)\\

\cite{freund2021implementation} & Offices building  & TAC and mechanical ventilation & 90 days & Nonlinear  & BMS-LON network (Not specified)\\

\cite{carli2020iot} & Laboratory & Fan coil & 4 months  & Linear  & IoT framework (Wireless) \\

\cite{hou2022model} & University building & Radiator and AHUs  & 7 days  & Nonlinear & NM\\

\cite{vivian2022experimental}  & Lightweight building lab & HVAC & 3 days & Linear & BEMS (Wired) \\
  \bottomrule                  
\label{expr}  
\end{tabular*}
\end{table*}

\section{Review of the Experimental Related Works}
\label{section:2}

The evaluation of energy-efficient HVAC systems incorporating MPC has typically been performed through simulations or physical hardware testing in real-world scenarios. In this section, we compiled a comprehensive list of published experimental studies, to the best of our knowledge, that utilized MPC in buildings between 2002 and 2022. Among these investigations, we have highlighted the most relevant. Table \ref{expr} summarizes these studies and compares them from various perspectives.

Several factors influence MPC setups in building control methods, including the type of actuators, communication methods, automation systems, and building model complexity. Similarly, the variety and complexity of mechanical systems have been controlled using MPC, such as a radiator, Air Handling Unit (AHU), ice storage, chiller, gas, electrical boiler, fan coil, windows and veils, floor heating, TABs, TACs, air conditioners, and cooling towers. Most studies control actuators using an existing low-level classic controller, such as PI, PID, or another embedded controller, in experimental devices or directly \cite{aswani2011reducing, carli2020iot}.

Most studies have adhered to a specific range of setpoints following ASHRAE \cite{de2002thermal} standards; however, incorporating resident feedback can significantly affect their comfort level, and user interactions with HVAC systems can improve their experience. Most studies consider that resident feedback can be gathered through direct or indirect methods. A few studies, such as \cite{schuss2011empirical, castilla2014thermal, aswani2011reducing,bengea2014implementation,de2016practical}, indirectly obtained feedback from occupants and considered the user input in some under-controlled devices. Occupancy-behavior detection \cite{dong2014real} is a direct example of gathering user feedback. In addition, a few studies \cite{schuss2011empirical,castilla2014thermal,fiorentini2017hybrid} used predicted mean vote (PMV) and PMV index to determine user comfort.

Experimental investigations primarily use MPC as a high-level regulator for Building Energy Management Systems (BEMS) or Building Automation Systems (BAS). This involves integrating sensing, control, monitoring, human-in-the-loop, and actuating subsystems. The communication platforms utilized in these experiments varied from wireless sensor networks to wired networks, depending on the specific building. Although research on MPC for regulating heating and cooling devices in various building types demonstrates impressive outcomes in terms of energy conservation and occupant satisfaction, these methods are limited to specific sites. This is because of the unique BAS or BEMS infrastructure and variations in modeling, which prevent generalization to other buildings.

An experimental study was conducted using an IoT framework to control the indoor temperature of a building using MPC control algorithms, which demonstrated the effectiveness of the IoT approach in the roles of BEMS and BAS. Researchers in \cite{carli2020iot} studied linear internal functions to maintain PMV constraints over four months during the warm season in an approximately 16-square-meter laboratory, where three sensor measurements were required.

After conducting a literature review of two approaches using MPC for building energy and occupant comfort management, IoT frameworks, and BEMS- and BAS-equipped building approaches, several research gaps were identified. There is a lack of experimental research on two digital \footnote{Digital controllers operate using binary options, with choices limited to either "ON" or "OFF."} or analog-controlled\footnote{Analog controllers have the capacity to process input within a range from 0 to 1 through the use of inverter drivers.} central AHU systems and ML produce a daily linear thermal model that serves as an internal MPC model, particularly in scenarios where no low-level controller exists in either approach.
There is a need for research on ML development to aid MPC when it only uses data from the building under MPC control, which is yet to be investigated \cite{yang2020model}.
Large-scale and long-term MPC experiments, in which an IoT framework employs a standalone BAS or BEMS that considers direct feedback from residents, remain to be implemented.

This study addresses this gap by presenting a cost-efficient, scalable, and adaptable IoT framework that can be used for buildings. The proposed framework used machine learning to generate a building model. MPC uses a model to regulate central air effectively using digital or analog controllers. Furthermore, equation-oriented controllers can also utilize the generated building model to enhance the performance of low-level controllers, such as PI or PID.

\section{IoT framework}
\label{section:3}
The following section discusses various aspects of the AHU system architecture, focusing on the proposed IoT framework.
The framework consists of the following components:

\begin{itemize}
    \item Sensor nodes and gateways: The nodes are in charge of collecting data from the environment and transmitting it to the gateways, which then aggregate the data and send it to the central server.
    
    \item Central server: It serves as a monitoring system for the Average Indoor Temperature (AIT) and power consumption, receives feedback from users or administrators, and enables automation control for the building.
    
    \item Actuators: They execute the generated decisions on the low-level controllers.
\end{itemize}
The overall architecture and its various components are illustrated in Figure \ref{IoT-solution}. Each component is described in detail in the following sections.

\subsection{Sensor nodes and gateways}
\begin{figure}[ht]
\centering
\includegraphics[width=0.395\textwidth]{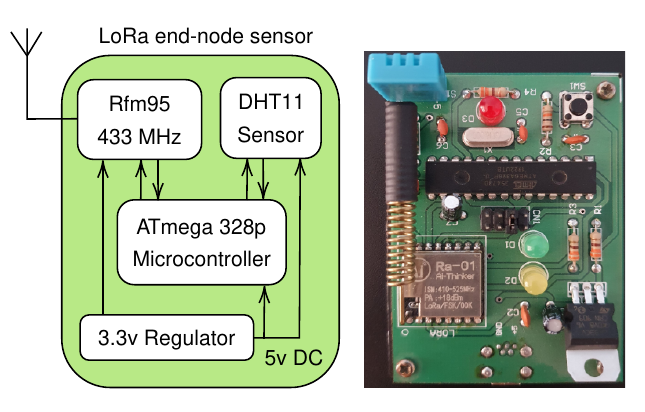}
\caption{The long-range sensor node was designed using a microcontroller and manufactured in this study.}
\label{lora-endnode}
\end{figure}

We designed the boards of each wireless sensor node to consume less energy when communicating via a wireless channel with the gateway. Figure \ref{lora-endnode} The LoRa end node depicts a detailed blueprint of the sensor nodes. The sensor nodes used the LoRa module for communication. In addition, we used an Atmega328 microcontroller to communicate with the LoRa SX1276 module and DHT11 sensor, which can measure the temperature and humidity with a precision of $\pm$2$ ^\circ C$ and $\pm$5\% RH, respectively, \cite{gay2018dht11,mutescu2021wireless}.

The sensor nodes capture and transmit data to the gateway every 5 min, whereas the gateway aggregates the data from different nodes. In our testbed, we used Raspberry Pi 3 devices as the gateways responsible for transmitting the aggregated data cyclically. The data were formatted in JSON, including the sensor ID, temperature, humidity, and date value, and transmitted using the MQTT protocol, which follows the publish-subscribe method. The central server periodically receives and processes data and stores them in a database for decision-making and monitoring purposes.

\subsection{Central Server}

The central server of the system consists of various components, including databases for recording the average temperature and humidity of all the sensors and MPC movements. It also includes a web server that provides a user or administrator interface for feedback, and a web server that offers a dashboard for monitoring the system in real time.

The data received by the gateways are stored in a NoSQL database, specifically MongoDB, which we refer to as the sensor database. In addition to the data, metadata were generated based on the information obtained. This includes the date of data reception and the average temperature and humidity of each floor, which were calculated by averaging all the received data every five minutes and saving it as a single value for both. The system also tracks connectivity problems with sensor nodes, and these issues are presented to the administrator for resolution.

\begin{figure*}[t]
    \centering
    {\includegraphics[width=8cm]{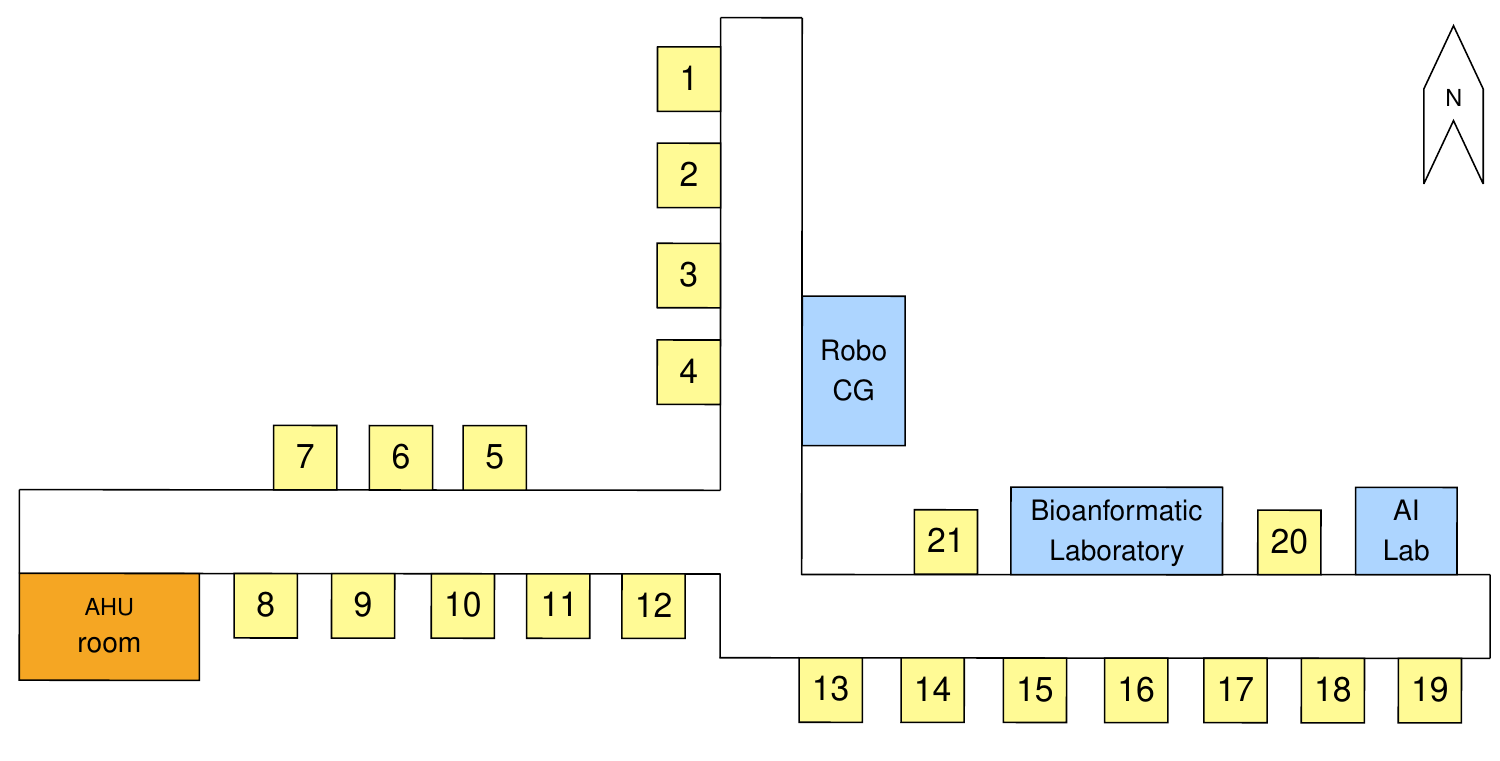} }
    \qquad
    {\includegraphics[width=8cm]{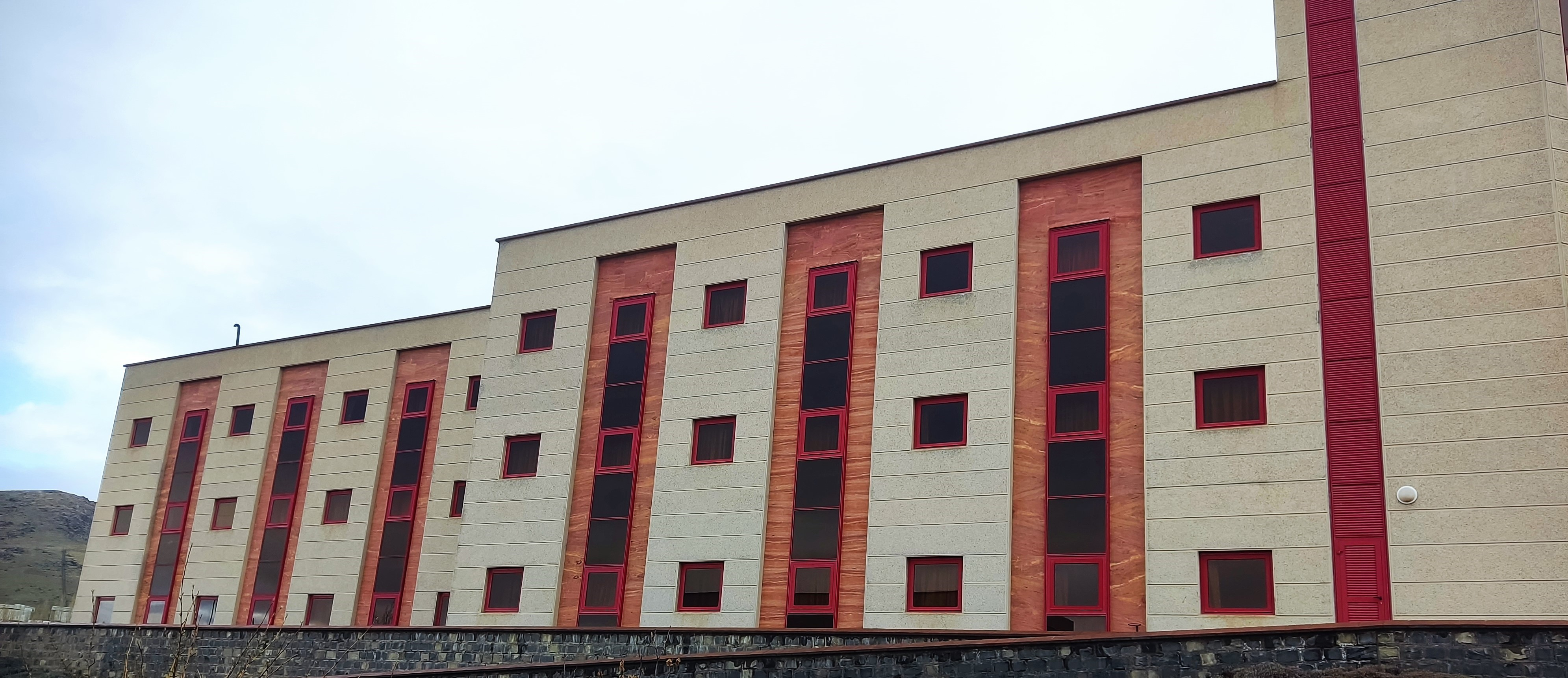} }

    \caption{Test environment overview (left) Blueprint of the building map indicating the locations of AHUs and rooms. (Right) Photograph of the building where the sunbeam shines on the southern side.}%
    \label{real and blueprint}
\end{figure*}

The central server is built using Node.js, and includes a web server that provides a user-friendly interface for both administrators and occupants to monitor data. Administrators can schedule AHU to operate on specific working days and times. The system has an additional interface for users to provide feedback on their desired room temperature. The system optimizes the decision-making process by utilizing an ML-based MPC that considers the setpoints of the occupants and administrators while minimizing the usage of the AHU. This MPC uses data stored or generated by the server, and is explained in more detail in the following sections. Occupants are not obliged to provide feedback, and can use the system only when they want to adjust the temperature.

\subsection{Actuator and testbed building}
A relay board was used to command the AHU connected to the main computer via a USB port. Wiring was conducted considering the minor interference in the AHU control system and switching to the manual control mode (clock-based controller) in the case of errors and failure of the proposed solution.
The studied building had three stories, and on each floor, two AHUs were placed on the northern and southern sides of the floor, in which facades were present in the southern offices. 

\begin{figure}[ht]
\centering
\includegraphics[width=0.40\textwidth]{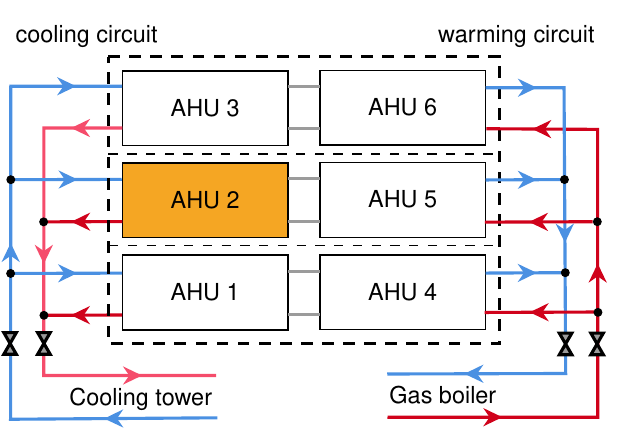}
\caption{The AHU equipment diagram displays the direction of the hot and cold water flow during the heating and cooling processes using blue and red arrows. Manual valves are used to prevent hot water from entering the system when the cooling tower is active and vice versa.}
\label{airwasher schematic}
\end{figure}

Figure \ref{real and blueprint} designated as AHU for control purposes, is located on the southern side of the second floor.
The occupants manually controlled all laboratories and room outlets. The AHUs are fed their coils with cold and hot water (Figure \ref{airwasher schematic}), producing filtered streams of hot and cold air using a three-phase electric fan controlled in winter and summer by turning the motor contactor ON or OFF. Additionally, two contactors for the water pump sprayed water in the AHU in the summer for 21 rooms and three laboratories through built-in metal ducts and entered the commissioned rooms’ outlets.

\section{Machine learning for thermal model}
\label{section:4}

\begin{figure}[ht]
\centering
\includegraphics[width=0.9\textwidth, trim = 2.5cm 19.1cm 4.5cm 4cm, clip]{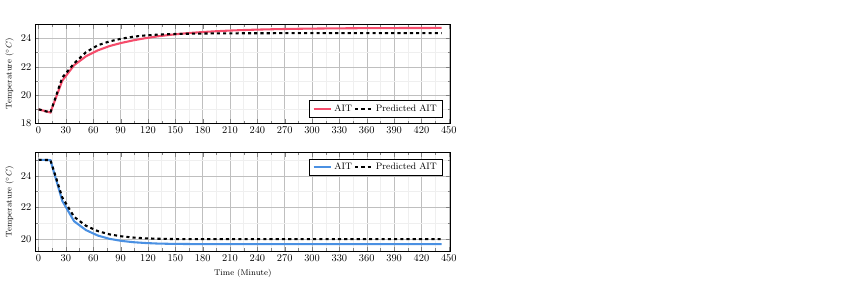}
\caption{Increasing (top) and decreasing (bottom) EDFs generated by ML and measured data for two random days between the previous year, mid-February, and mid-March.}
\label{real and predicted curves}
\end{figure}

The building thermal model is essential for control algorithms, and this section employs machine learning to present a daily adaptive-linear thermal model. This study combined the AHU system, building model, and disturbance models to develop a thermal model. Although there is a nonlinear relationship among the factors affecting the building's thermal properties, the data collected from the WSN indicates that the performance of the AHU on the AIT of the building is comparable to a first-order dynamic system, as demonstrated in Figure \ref{real and predicted curves}. Simplifying the model makes it easier to implement without causing a significant drop in accuracy. \cite{boodi2022building}.

\begin{figure*}[t]
    \centering

    {\includegraphics[width=0.47\textwidth, trim = 5cm 18cm 4.8cm 4cm, clip]
    {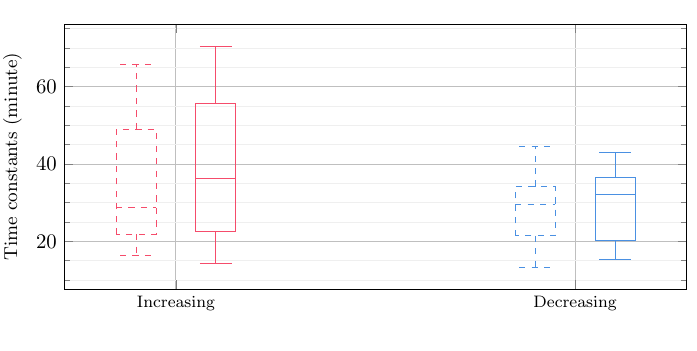} }%
    \qquad
    {\includegraphics[width=0.47\textwidth, trim = 5cm 18cm 4.8cm 4cm, clip]{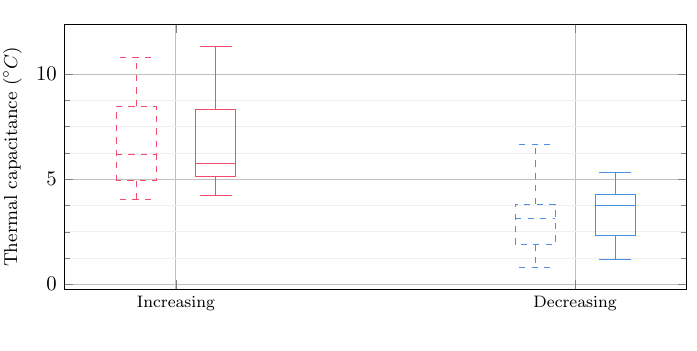} }%
    \caption{
A comparison between the measured (solid) and predicted (dashed) gain and time constant parameters for the increasing and decreasing trends was conducted over one month, from mid-February to mid-March. The right panel displays the gain parameters and the left panel shows the time constants.}
    \label{predicted and measured gain and time constants}
\end{figure*}

The equation describing the thermal dynamics in the time domain, accounting for the delay ($\theta$), is obtained by performing an inverse Laplace transform on the step input ($1/s$) response of the first-order system (FOS) transfer function, denoted by $G(s)$ which represents the thermal dynamics in the complex frequency domain $s$. This results in a general thermal equation with delay:

{\small\begin{align}
  \begin{split}
    &
      G(s) = \frac {k_p}{\tau s + 1}.e^{-\theta s}
     \\&
     y(t < \theta) = y_{init}
     \\&
     y(t\geq\theta) = u(t-\theta).k_{p}(1 - e^{-(t-\theta)/\tau}) + y_{init}(e^{-(t-\theta)/\tau}).
      \\
   \end{split}
   \label{func}
\end{align}}

AIT is represented by the variable $y(t)$ in this system, where $y_{init}$ denotes the initial temperature. The time taken by the AIT to reach approximately 63\% of its thermal capacitance is represented by the time constant $\tau$, which is denoted by the gain, $k_p$.
Input $u(t)$ represents the step function applied to the system, with a gain ranging from 0 to 1. The delay was assumed to be constant at 13 min based on the experimental data from our testbed.
The sign of the gain parameter ($k_p$) in descending processes, such as when the AHU operation stops and the building's AIT begins to decrease, is negative. Overall, there were no signs of thermal capacitance.

As the building model is subject to daily changes owing to disturbances, the generated curves for increasing and decreasing AIT during a one-month learning period (mid-February and mid-March) yield varying $k_p$ and $\tau$, along with the gain and time constants of the measured parameters, as shown in Figure \ref{predicted and measured gain and time constants}. The primary objective of this technique is to obtain finely tuned model parameters that can be utilized in equation-oriented control algorithms.

To obtain the model parameters, it is necessary to have daily access to environmental dynamic function (EDF) data. The EDF provides the gain and time constant parameters for the FOS model used in MPC. The FOS gain ($k_p$) parameter was calculated by determining the maximum and minimum values of the measured or EDF curves, and the time constants ($\tau$) were obtained by determining the time at which the measured or EDF curves reached 98\% of the thermal capacitance ($4\tau$) \cite{Connell2017}.

This approach simplifies the MPC process by avoiding the need to model disturbances in the optimization problem. In addition, an Artificial Neural Network (ANN) was used to learn the EDF and predict the AIT over the desired horizons and time intervals, making this technique applicable to both warm and cold seasons.

The ANN provides a mapping from the input data, including the initial AIT ($T_{init}$), time interval to the next AIT ($\Delta t$), AHU input value ($I_{AHU}$), outdoor temperature ($T_{out}$), humidity ($H_{out}$), wind speed ($W_{speed}$), solar radiation ($S_{rad}$), and energy ($S_{energy}$), to the output data, which is the difference in temperature between the AIT after the time interval and the initial AIT.

\begin{align}
  \begin{split}
    &
     Input: < T_{init}, \Delta t, I_{AHU}, T_{out}, H_{out}, W_{speed}, S_{rad}, S_{energy}>
     \\&
      Output: < T_{\Delta t} - T_{init}>
      \\
   \end{split}
   \label{ML I/O form}
\end{align}

\begin{figure}[ht]
\centering
\includegraphics[width=0.39\textwidth]{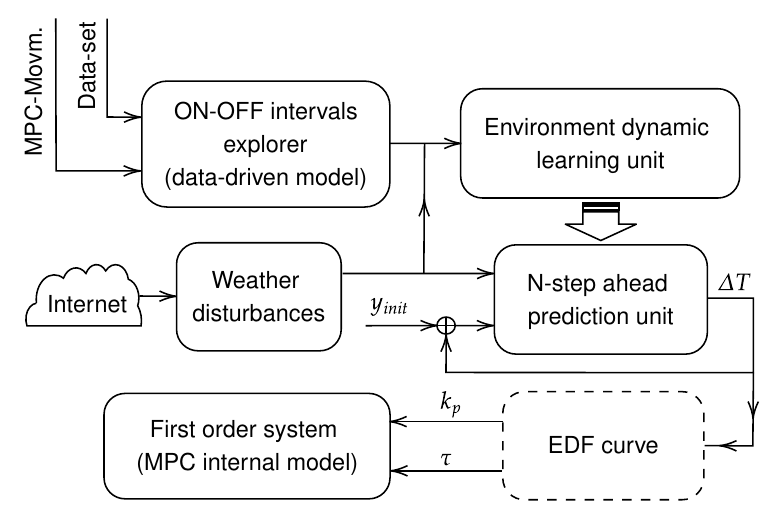}
\caption{The ML diagram presents hierarchical data preprocessing for EDFs learning and prediction, extracting time constants and gain from EDFs generated by disturbances and temperature gradients.}
\label{ML-diagram}
\end{figure}

Figure \ref{ML-diagram} illustrates that by training the model with samples gathered from the AIT, AHU input, and weather information measurements in the defined form (notation \ref{ML I/O form}) as disturbances, the sequential prediction technique can generate increasing and decreasing dayahead EDF curves. These predictions rely on the initial AIT, the time interval, the maximum input gain, and disturbances.

To accomplish this goal, a one-month dataset from the Previous Year (PY) was employed to train the EDF model, which was then evaluated against the measured data from the same period. It is worth mentioning that the measured AIT of the PY consisted of continuous data from prolonged periods when the AHU was operating at its maximum input gain. Therefore, the $\tau$ and $k_p$ parameters obtained from the measured data were available for comparison with those of the ANN model.

\begin{table*}
\caption{Evaluation of k-fold CV-predicted temperature gradient on test data for various metrics (increasing/decreasing).}
 \begin{tabular*}{\textwidth}{l @{\extracolsep{\fill}}llllll}
  \toprule
  \textbf{Months} & \textbf{Dataset size} & \textbf{MSE} & \textbf{Scaled MAE} & \textbf{Explained variance} & \textbf{R squared} \\ 
  \midrule
Mid-Feb to Mid-March (PY)   & 79550 / 240080  & 0.0039 / 0.0034 & 0.1045 / 0.0978 & 0.8546 / 0.6122   & 0.8283 / 0.5991   \\  \midrule

November &  65122 / 102600  & 0.0043 / 0.0066 &  0.1026 / 0.0911 & 0.7901 / 0.6302 & 0.7445 / 0.6135 \\

December    &  106587 / 167151  & 0.0045 / 0.0065 & 0.0937 / 0.0792 & 0.7513 / 0.6161 & 0.7310 / 0.6015 \\

January   &  107023 / 165341 & 0.0033 / 0.0046  & 0.0904 / 0.1061 & 0.7629 / 0.6399 & 0.7375 / 0.6139\\

February   & 97536 / 153269 & 0.0030 / 0.0053  &  0.0975 / 0.0886 & 0.7351 / 0.6298 & 0.7165 / 0.6108\\

March    & 54281 / 88991  &   0.0038 / 0.0058 &  0.1139 / 0.0780 & 0.7813 / 0.6684 & 0.7595 / 0.6144\\

Entire experiment duration    & 430549 / 677352 &  0.0031 / 0.0059  & 0.0925 / 0.0687 & 0.7360 / 0.6262  & 0.7222 / 0.6034\\

\bottomrule  
\label{learning-info}
\end{tabular*}
\end{table*}

To avoid overfitting, the model was trained using k-fold cross-validation (CV) techniques, utilizing five hidden layers, and tuning the hyperparameters through testing and trial runs. Table \ref{learning-info} provides both monthly and collective assessments of the trained ANN performance with unseen data. Each dataset was split into three subsets, with 70\% allocated for training, 15\% for validation, and 15\% for testing the unseen data.
 
The performance of the model is highlighted by its low Mean Squared Error (MSE) and scaled Mean Absolute Error (MAE) values, which indicate that the model can make highly accurate predictions with minimal errors. Furthermore, the model exhibits a high explained variance and $R^2$ score, implying that the trained model can account for a substantial proportion of the variance in the target variable.

\begin{figure}[ht]
\centering
\includegraphics[width=0.38\textwidth]{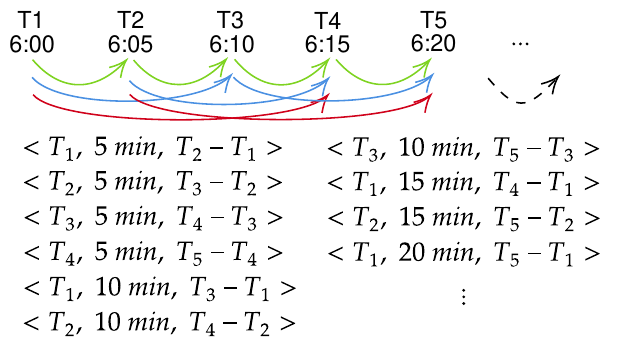}
\caption{Example of providing samples from existing samples for a session to create a daily data-driven model.}
\label{ti-interval-diagram}
\end{figure}

The data-driven model was developed by analyzing the AHU operating sessions during the real-time MPC experiment and AIT at various time resolutions ranging from 5 to 300 min. Two data-driven models were created to increase and decrease the building AIT. As shown in Figure \ref{ti-interval-diagram}, this was accomplished by producing all two-member records within the validated time interval of the AHU operating sessions in a dataset.

The system employs two databases: MPC-Movements and Dataset. The former logs all the movements of the MPC algorithm, including the AIT, setpoints, optimal control actions, and record dates. The dataset (NoSQL DB) stores average data from the installed sensors. AIT sessions can be retrieved by accessing the MPC movement database. Weather data downloaded from the Internet \cite{weather} were integrated into the data-driven model as the ML input using a daily timeframe (notation \ref{ML I/O form}).

Because the underlying patterns in the dataset change slowly over time, increasing the sample number at various intervals could assist in covering the noise in the generated EDFs, such as changes in the occupancy level, window open or closed status, and momentary changes in the AHU input pipe water temperature, or irregular office activities. Additionally, there are cases in which no high-resolution sensor data sampling is available to obtain the daily EDF \cite{s17061221}. These techniques bring us about 3400 and 5400 data daily for obtaining building thermal models for increasing and decreasing.

\begin{figure}[t]
\centering
\includegraphics[width=0.53\textwidth,height=0.41\textwidth, trim = 5cm 13.7cm 3cm 4cm, clip]{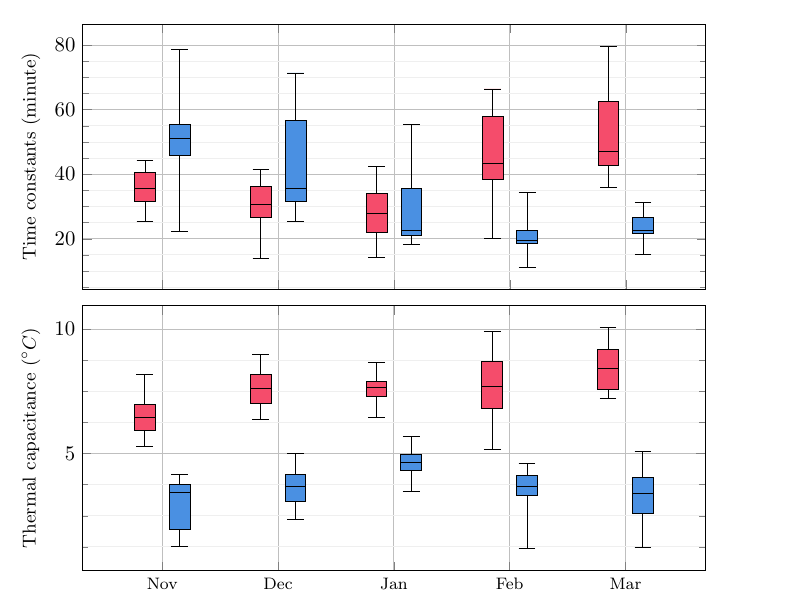}
\caption{Comparison between monthly predicted time constants (top) and gains (bottom) of EDFs, both increasing (red) and decreasing (blue) for the entire duration of implementation.}
\label{5-month gain}
\end{figure}

Figure \ref{5-month gain} depicts the FOS parameters derived from the EDFs, which serve as an internal model in MPC and for nonlinear output mapping for each month. It also demonstrates the correlation between the month and thermal capacitance for increasing and decreasing AIT, revealing a trade-off between the time constants for warming up and cooling down the building. It is worth noting that this tradeoff shifts in mid-January, with the cooling time constants becoming considerably less than warming up because of the cold outdoor temperature.

\section{MPC Formulation}
\label{section:5}

To run the MPC algorithm in real-time and make decisions for AHU based on the given setpoints and WSN data, the following routine will be conducted every day: \\

1) At 6 a.m., the trained model will generate increasing and decreasing EDFs based on notation \ref{ML I/O form} and Figure \ref{ML-diagram}, and obtain FOSs parameters $\tau$ and $k_p$. It used the latest AIT, maximum input gain, updated disturbances, and defined time resolution, which considered equal delay parameters of 13 minutes. These parameters were used in the MPC internal model and nonlinear output mapping.

2) When the MPC enters idle mode, the setpoints are reduced below the AIT from 9 p.m. to 6 a.m. During the idle period between 12:30 a.m. and 5:30 a.m., the framework generates yesterday's (6 a.m. to 12 a.m.) data-driven model using MPC movements and AIT data and integrates it with the disturbances of yesterday. This model was then appended to the dataset from the previous year or the previous day. The model is then trained using this information. It is worth noting that the training dataset window length was limited to two consecutive months.

\begin{figure}[h]
\includegraphics[width=.48\textwidth]{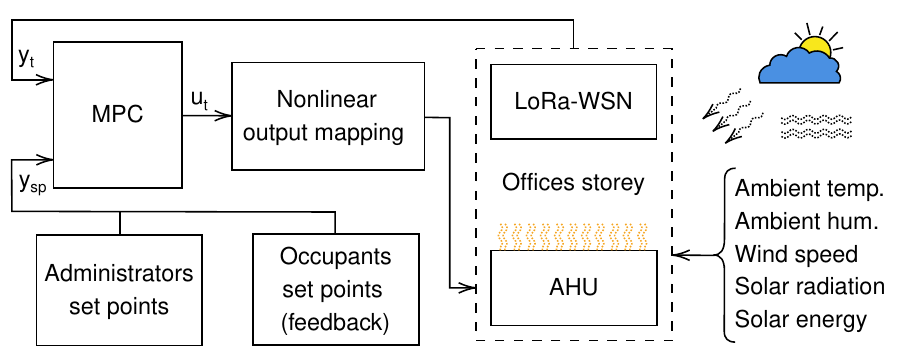}
\caption{This figure illustrates the closed-loop architecture of the MPC system. Data from the WSN and feedback (occupants or administrators) are presented to produce the optimal input for the AHU based on the buildings' disturbances.}
\label{MPC-diagram}
\end{figure}

The FOS parameters obtained from the EDF increasing curve were used in the MPC as an internal model. The prediction horizon was set to 48 (as well as the control horizon), denoted as $p$, the sampling time was 30 min, and the execution time was 24 h. The problem is defined in the form of a minimization problem (Equation \ref{mpc-func}), where the inputs are the numerical average of setpoint feedbacks and the AIT, and the output is the float number declaring the optimal computed movement for the AHU as illustrated in Figure \ref{MPC-diagram}. 

\begin{align}
  \begin{split}
     & \min_{U_k}\{\sum_{k=0}^{p-1} (\lVert Y_{k} - Y_{sp}\rVert ^2 + \lVert \Delta{U_{k}} \rVert ^2) \} \\ &
     s.t.: 
     x_{k+1} = f(x_{k},u_{k})
     \\ &
     y_{k} = g(x_{k})
     \\ &
     u_{k} \in [0,1]
     \\ &
     x_k = x(t)
     \\ &
     k = 0, 1, ..., p-1 .
   \end{split}
   \label{mpc-func}
\end{align}

At each sampling time (each cycle of calculation) of the MPC algorithm, $x_{k+1} = f(x_{k},u_{k})$ and  $y_{k} = g(x_{k})$ are the state spaces of the system thermal model, which is taken from the FOSs. $Y_k$ is defined as the AIT vector of the next AITs over the prediction horizon, as well as $Y_{sp}$ is the setpoint vector, and $\Delta{U_{k}}$, is the difference vector of the next produced inputs. $y_{sp}$ is the setpoint temperature from the administrators' setpoints or feedback from the occupants, and $u_k$ is the control output.

\begin{figure}[ht]
\centering
\includegraphics[width=0.69\textwidth, trim = 5.2cm 18.2cm 2cm 4cm, clip]{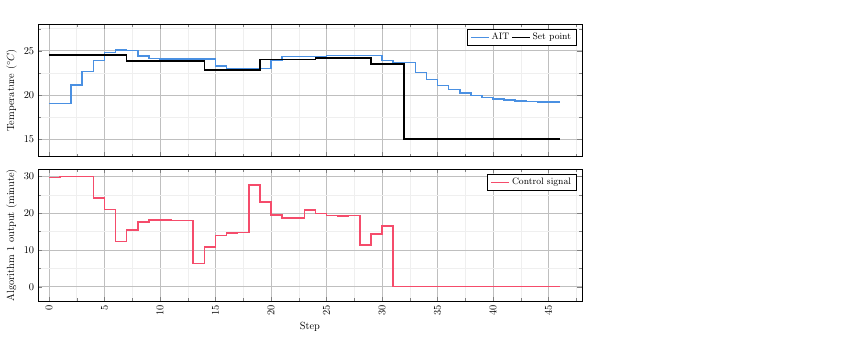}
\caption{Graphs of the MPC performance for tracking setpoints in a day (top) and the output of Algorithm \ref{preprocessing-software}, which mapped the optimal MPC control signal (bottom).}
\label{mpc-one-day}
\end{figure}

The objective function is also defined as the difference between the setpoints and AIT and the differences between the produced AHU inputs. For each sampling time, the collected average setpoints from occupants and the latest AIT are presented for MPC calculation to minimize the objective function by computing ${U_k}$, which is defined as the optimal control vector over the determined prediction horizon for the AHU that optimally tracks setpoints (Figure \ref{mpc-one-day}). 

The control output constraint was also considered to support analog AHU systems. If feedback from the occupants or administrators is unavailable, the program sets the temperature within the range of 20-25$^\circ C$. The feedback is valid within a time interval of $4\tau$, enabling the AHU to respond to the given setpoint. A degree and thirty-three hundredths greater than the ASHRAE standard.

Restriction of the AHU control system (binary control) leads to the consideration of a nonlinear function to map control actions between 0 and 1 in a defined sampling interval time to variable time (less than or equal to the sampling interval time) in a constant control action ON or OFF \cite{american2008building}.
\begin{algorithm}
\algsetup{linenosize=\small}

\caption{Nonlinear output mapping}
\label{preprocessing-software}
\begin{algorithmic}[1]
\STATE Input: $u_k$, $FOS_{increasing}$, $FOS_{decreasing}$
\STATE Output: The optimal AHU ON time ($t^{*}$) 
\STATE {$T_{init} \leftarrow Latest \ AIT $}
\FOR{ $t:1~to~30$}
\STATE $T_{mid}\leftarrow FOS_{increasing}(u(t)=1, \textit{time}=t , y_{init}=T_{init})$
\STATE $T_{end}\leftarrow FOS_{decreasing}(u(t)=1, \textit{time}=t_{sampling} - t , y_{init}=T_{mid})$ 
\IF {$|FOS_{increasing}(u(t)=u_k ,\textit{time}=t_{sampling} ,y_{init}=T_{init})-T_{end}|\leqslant \epsilon $}
\STATE $t^{*} \leftarrow t$
\STATE Go to line 12
\ENDIF
\ENDFOR
\STATE Return $t^{*}$
\end{algorithmic}
\end{algorithm}

Note that for the analog-controlled AHU, the need for mapping is eliminated, and the MPC-produced control signal can be directly connected to the driver. Algorithm \ref{preprocessing-software} describes the generation of the optimized duration to turn ON the AHU to achieve the next-ahead predicted temperature using MPC. Based on FOSs, a brute-force algorithm determines the precise duration. $t_{sampling}$ is considered as the sampling time (30 min), and $t^{*}$ is the outcome of Algorithm \ref{preprocessing-software}.

An optional protection parameter can be introduced to protect the electric motor from short ON durations. There are two possible scenarios for this.

1) If $t^{*}$ is less than or equal to the protection parameter, then round $t^{*}$ is zero. This helps to protect the motor against a short duration ON time while very slightly losing the temperature

2) If the difference between the sampling time and $t^{*}$ is less than or equal to the protection parameters, the motor is kept ON, corresponding to the defined sampling time.

\section{Results}
\label{section:6}

This section analyzes the performance of the IoT framework on existing AHU, focusing on several key factors: AIT, occupant comfort, energy consumption, energy savings compared to manual control, and AHU operation during the two warm months in summer. The proposed method was conducted for over 126 days, which is almost the coldest days of the year, from November 11 to March 16.

\begin{figure}[ht]
\centering
\includegraphics[width=1.05\textwidth, trim = 2.5cm 21.0cm 2cm 4cm, clip]{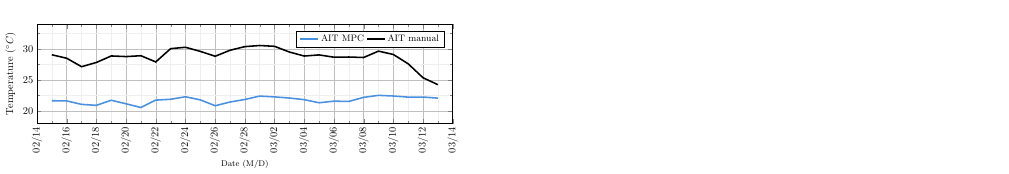}
\caption{The comparison of the AIT between the proposed IoT framework (blue) in the same period of two consecutive years in the testbed building and manual control (black) in which proof of manual control inefficiency.}
\label{consecutive years}
\end{figure} 

Figure \ref{consecutive years} shows that the AIT of the proposed method is consistently lower than that of manual control during two consecutive years, with fluctuations ranging between 20 and 22.5$ ^\circ C$. It was also shown that the ML unit, in providing the daily thermal model, achieved good results in MPC to maintain the AIT in the desired range defined by administrators. The black curve in the chart indicates high temperatures and explains the energy wastage caused by occupants opening their office windows. It is worth noting that sudden temperature drops may be attributed to changes in the clock timer control made by technicians due to the low activities of departments during the year-end holidays.

\subsection{Temperature analyzing and occupants' comfort}

\begin{figure}[t]
\centering
\includegraphics[width=1\textwidth, trim = 2.5cm 17cm 2cm 4cm, clip]{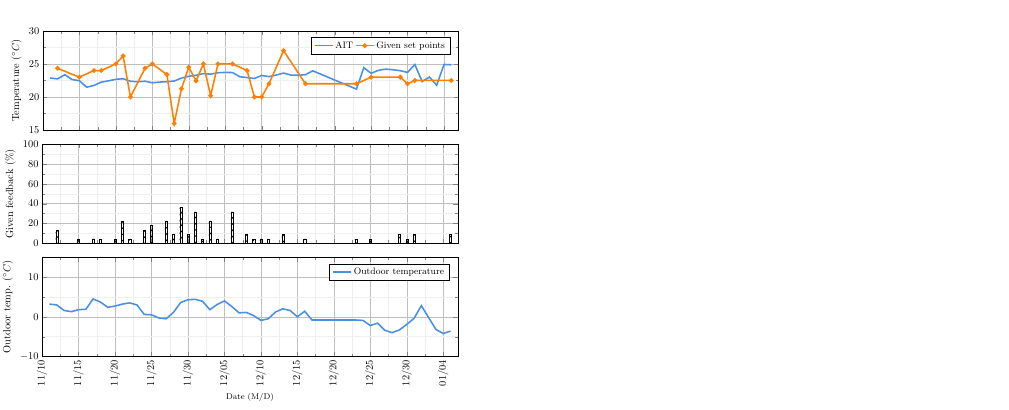}

\caption{The graphs of the outdoor temperature (bottom), the percentage of occupants (center), and the average given temperature from occupants and daily AIT (top) from 11 November to 5 January.}
\label{F1}
\end{figure}

\begin{figure}[t]
\centering
\includegraphics[width=1\textwidth, trim = 2.5cm 17cm 2cm 4cm, clip]{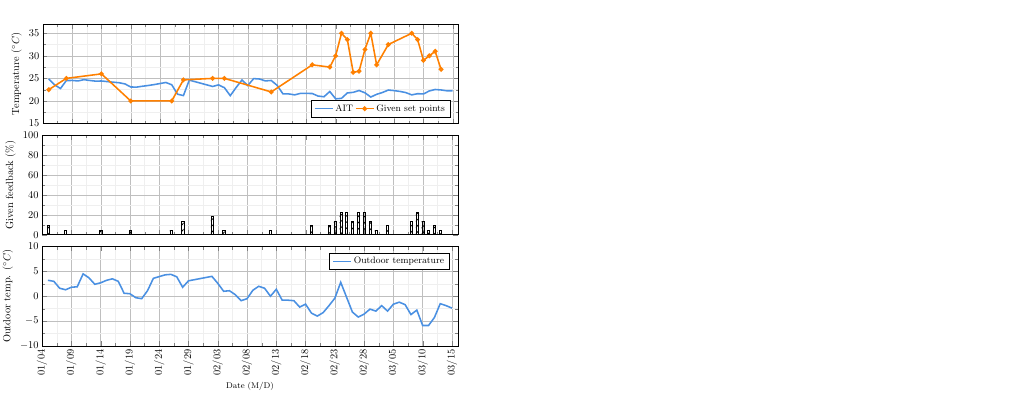}
\caption{The graphs depict the average temperature given by occupants, daily AIT (top), percentage of participation by occupants (center), and outdoor temperature (bottom) from January 5th to March 15th. The high percentage of participation was due to the administrator setting the temperature lower because of the institute's low activities on the Nowruz holiday.}
\label{F2}
\end{figure}

Figure \ref{F1} and \ref{F2} display graphs of the feedback received from users who registered on the server. During the implementation of the proposed IoT solution, 140 feedback points were recorded. 

It is worth noting that the highest percentage of numerical feedback received in a single day was 38\%, whereas the average number of feedback reports received daily was approximately 1.1. This indicated that the average feedback percentage for a single day was 4.6\%. During the university's busy season, using a web interface in 21 rooms and three laboratories could be a positive indicator of user satisfaction. 

The web UI simplifies the temperature control process without requiring complex manual adjustments, offering a convenient and accessible solution. It is important to mention that users can input unusually high or low setpoint values. These unusual setpoints, referred to as 'emotional feedback,' present a challenge that can be addressed by disregarding these outlier values.

\subsection{Energy consumption}

\begin{figure}[h]
\centering
\includegraphics[width=0.51\textwidth, trim = 2.5cm 18.1cm 2cm 4.5cm, clip]{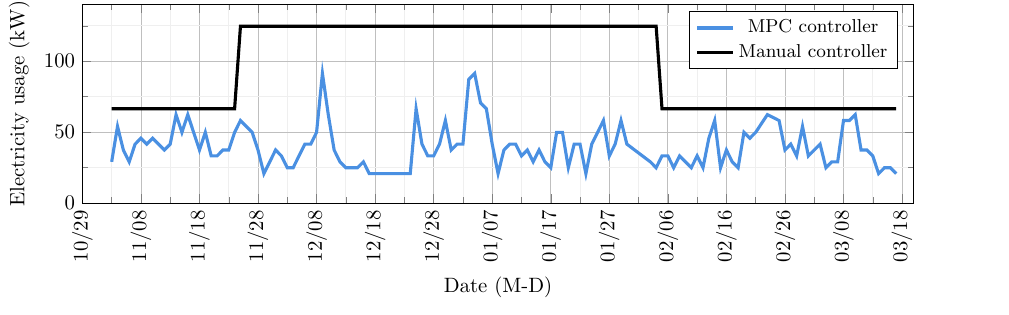}
\caption{Comparison of electric energy consumption between manual control and IoT solutions.}
\label{manual-mpc-energy}
\end{figure}

Figure \ref{manual-mpc-energy} displays the energy consumption of MPC control and manual control during the experimental period. Additionally, it demonstrates that in the proposed IoT framework, the energy consumption remains below that of manual control. Additional energy savings can be achieved by reducing the temperature difference between the input and output pipes, which can be achieved by using a lower AHU and decreasing the demand for the gas boiler.

The electrical energy consumption of a 3-phase electric motor was calculated using the following relation \cite{bose1986power}:
\begin{align}
  \begin{split}
  & P = U\times I \times Cos \phi \times \sqrt{3},
     \\
   \end{split}
   \label{energy-cons}
\end{align}

where $U$, $I$, and $Cos \phi$ are assumed to be 380 volts, 15.4 amperes, and 0.82, respectively. Thus, $P$ was approximately 8.3 $kWh$. From December 24 to January 30, the manual timer was set to start at 6 a.m. to approximately 9 p.m.; on two periods, November 11 to December 23 and January 31 to March 16, the manual timer was set for 8 hours consecutively.

\begin{figure}[ht]
\centering
\includegraphics[width=0.55\textwidth, trim = 4cm 16.5cm 2cm 4cm, clip]{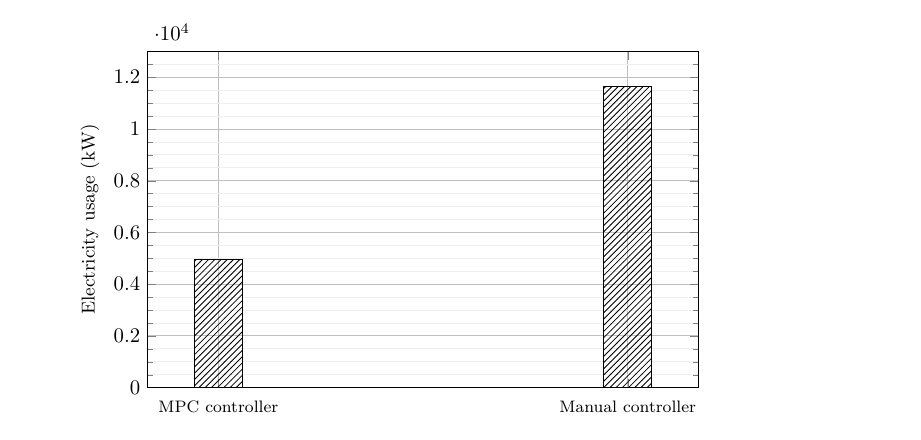}
\caption{The figure demonstrates the effectiveness of the IoT solution by comparing the total energy consumption of the manual controller and MPC controller, revealing a significant 57.59\% reduction in energy usage.}
\label{bar-chart}
\end{figure}

According to the data presented in Figure \ref{bar-chart}, the proposed method yielded a significantly lower electrical energy consumption of approximately 4920 kW compared with manual control, which consumed approximately 11660 kW (relation \ref{energy-cons}).

These results suggest a substantial energy saving of approximately 57.59\% when employing the IoT framework within the building.

\subsection{Performance of the AHU in the warm season}
\begin{figure}[ht]
\centering
\includegraphics[width=1.05\textwidth, trim = 2.5cm 21.0cm 2cm 4cm, clip]{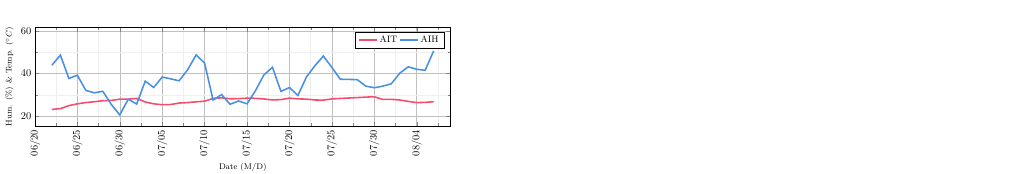}
\caption{In the summer, the performance graph of the AHU system displays a consistent trend, suggesting that the system may not operate properly.}
\label{summer}
\end{figure}

Throughout the summer season, from June 22 to August 7, the AHU system proved ineffective at lowering indoor temperatures, leading to an elevation in humidity levels. This observation was corroborated by the data presented in Figure \ref{summer}, which shows a consistently high average indoor humidity throughout the summer months, despite the continuous operation of the AHU. The occupants' feedback obtained through questionnaires indicated that they experienced discomfort and suffered from high humidity levels, attributable to the AHU's water nozzles releasing excessive humidity rather than decreasing air temperature. These findings suggest that the AHU cooling capacity was inadequate to reduce AIT.

The conclusion is that control algorithms would likely be ineffective during the summer unless additional measures were implemented to cool the input water pipe of the AHU, such as utilizing cooling towers or integrating an alternative device within the AHU itself.

\section{Conclusion and Future Work}
\label{section:7}
This study proposes a cost-efficient, scalable, and adaptable IoT framework for buildings with only temperature sensors and central air conditioners with digital or analog or equation-oriented controllers, including MPC and PI or PID. The primary goal is to demonstrate how IoT solutions can effectively enhance the efficiency of legacy HVAC systems, even in scenarios in which building information is limited. The proposed framework fills the gaps in the reviewed literature by presenting an ML-based MPC-IoT framework designed for digital or analog-controlled central AHU systems.

We provide extensive measurement results for the occupants' set-point tracking, accuracy of the ML-produced EDFs, daily energy consumption, and system performance. The results demonstrated that the proposed IoT solution can maintain user comfort and improve the performance of legacy systems without significant infrastructure upgrades. This further reveals that by utilizing stream data alone, ML can effectively produce a building model when employed with MPC as the sole controller.

This study shows the potential of a cost-effective solution for optimizing buildings with legacy HVAC systems. The implications of the proposed method go beyond reducing energy consumption. It can also have a positive impact on climate change by reducing gas emissions.

By exploring future avenues and extending our current research, we can delve into various aspects. The accuracy of the ANN thermal model can be enhanced by utilizing hourly weather disturbances, thereby improving the MPC model. To further improve the usage of WSN, several steps can be considered.

\begin{itemize}
\item Firstly, gathering additional information about the building, such as the temperature of AHU input and output pipes, building outside temperature, and humidity, can cover more nonlinearity of the thermal model and improve prediction accuracy. This provides valuable insights into a more accurate building model.
\item Secondly, using more accurate sensors than the DHT11 can improve the data quality. In addition, the framework uses an hourly timeframe to integrate weather disturbances with a data-driven model when dealing with diverse dataset samples.
\item Thirdly, utilizing energy harvesting methods from the environment can eliminate the need for sensor node batteries or wire adaptors. 
\end{itemize}

Furthermore, expanding the solution to cover the remaining five AHU units on building floors can save more energy. In addition, developing mechanisms to address users' emotional feedback and occupancy patterns can further improve user satisfaction and engagement. Moreover, a cloud-based IoT solution can provide more flexibility in its implementation.

\section*{Acknowledgment}
We have utilized online AI resources, OpenAI's ChatGPT~\cite{chatgpt} and PaperPal~\cite{paperpal}, to enhance the manuscript's English and to conduct grammar checks.

\bibliography{references}

\begin{IEEEbiography}
[{\includegraphics[width=1in,height=1.25in,clip,keepaspectratio]{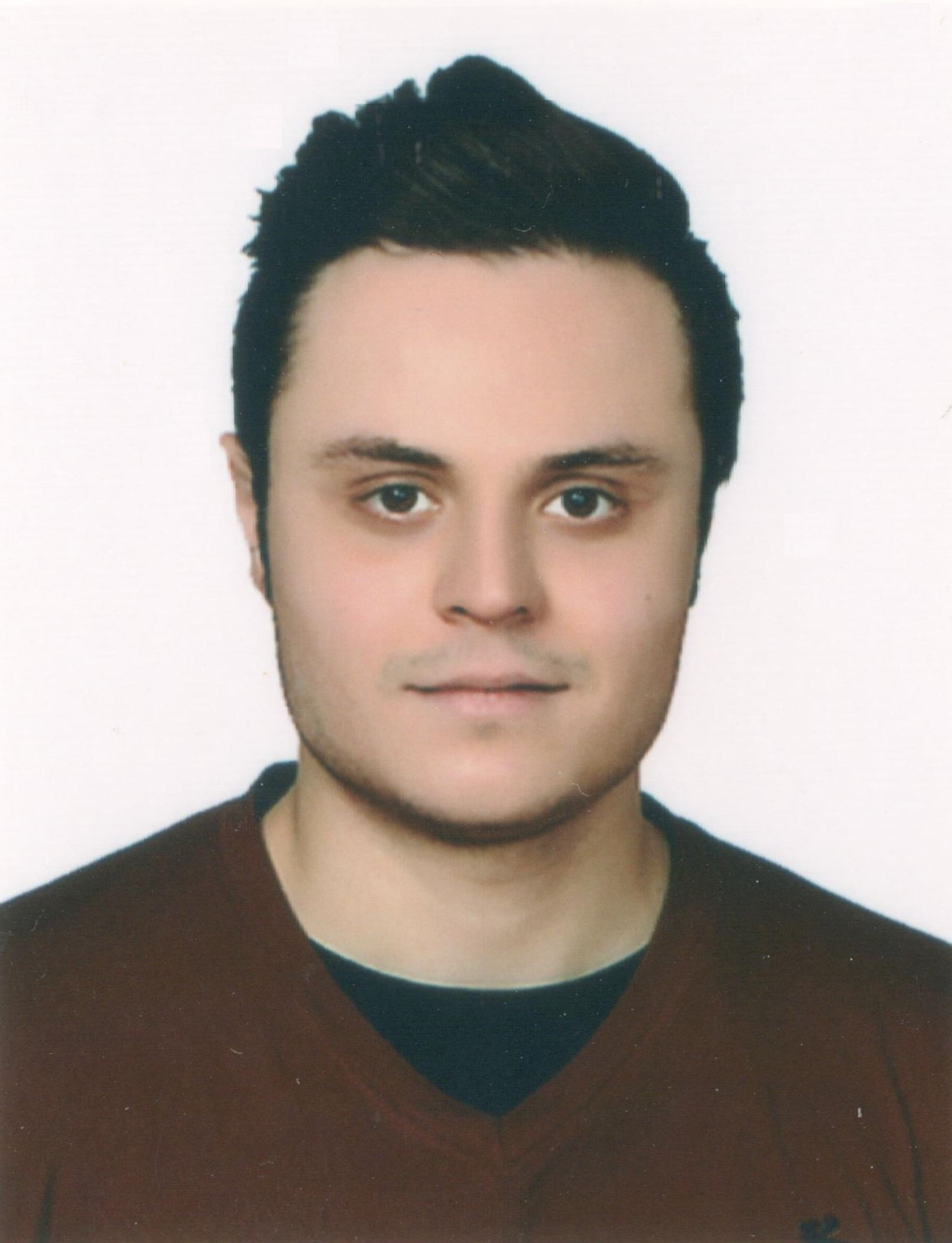}}]{Aryan Morteza} received the B.S. degrees in computer science at the University of Tabriz, in 2015, and the M.S. degree in computer science (smart systems) from the Institute for Advanced Studies in Basic Sciences (IASBS) in 2019.
From 2017 to 2019, he was a research intern in wireless communication and developing communication testbeds for IoT applications focusing on developing an IoT system for controlling HVAC based on the data-driven model and designing wireless sensor nodes for different communication protocols. His research interests include IoT and ICT solutions, artificial intelligence and machine learning, model predictive control, and wireless sensor networks.
\end{IEEEbiography}

\begin{IEEEbiography}[{\includegraphics[width=1in,height=1.25in, clip,keepaspectratio]{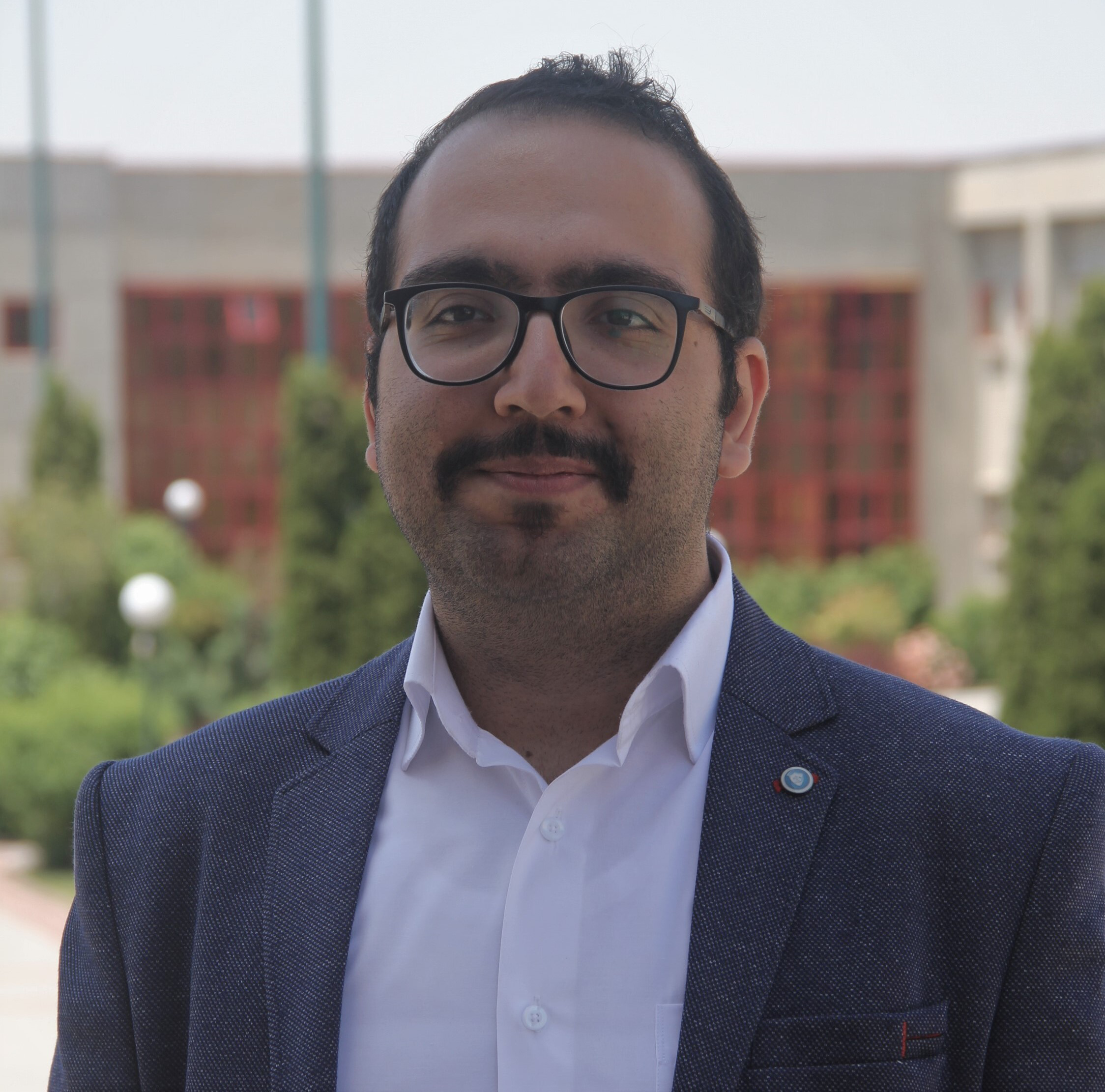}}]{Hosein K. Nazari} received his B.Sc. in Information Technology Engineering in 2018 from Institute for Advanced Studies in Basic Sciences (IASBS) in Zanjan, Iran. In 2021, he graduated with an M.Sc. in Computer Science from IASBS. He is currently pursuing a Ph.D. in Electrical and Computer Engineering Department of Technische Universität Dresden, Germany. His research interests include Network coding, IoT, and Time-Sensitive Networking.
\end{IEEEbiography}

\begin{IEEEbiography}[{\includegraphics[width=1in,height=1.25in,clip,keepaspectratio]{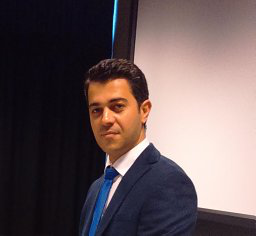}}]{Peyman Pahlevani}  is an Assistant Professor in the department of
Computer Science and Information Technology of the Institute for
Advanced Studies in Basic Science (IASBS), Iran. He received his
Ph.D. in Wireless Communication from Aalborg University, Denmark, in 2014. Prof. Pahlevani was a visiting researcher at the Computer Science Department of the University of California, Los Angeles (UCLA). Moreover, He has collaborated with different institutes and universities, such as MIT and Porto University. Besides his academic background, he extended his skills to solve practical challenges in the area of video communication in his previous career as a Researcher Engineer at AIRTAME company. His research interests are in wireless communication, network coding and its applications, vehicular communications, cooperative networking, and video streaming over WiFi links. Dr. Pahlevani has also served as a TPC member for international conferences and as a reviewer for high-impact journals, such as IEEE Transactions on Vehicular Technology.
\end{IEEEbiography}

\end{document}